\documentclass{ifacmtg} 
\usepackage{amsmath} 
\usepackage{amssymb}  
\usepackage[latin1]{inputenc}
\usepackage{graphicx}
\usepackage{pst-all}
\usepackage{subfigure}

\usepackage{theorem}
\newtheorem{defin}{Definition}

\usepackage{natbib}

\renewcommand{\matrix}[1]{\begin{pmatrix}#1\end{pmatrix}}
\newcommand{\T}{^\mathrm{T}}
\newcommand{\Real}{\mathbb{R}}

\newcommand{\unit}[1]{\mathrm{#1}}

\begin{document}
\runauthor{Waldherr, Eißing, Chaves and Allgöwer}
\begin{frontmatter}
\title{Bistability preserving model reduction in apoptosis}
\author{Steffen Waldherr\thanksref{mail}},
\author{Thomas Eißing},
\author{Madalena Chaves}, \textbf{and}
\author{Frank Allgöwer}
\address{Institute for Systems Theory and Automatic Control\\
University of Stuttgart\\
Pfaffenwaldring 9,
Stuttgart, Germany}
\thanks[mail]{Corresponding author (\texttt{waldherr@ist.uni-stuttgart.de})}

\begin{abstract}
Biological systems are typically very complex and need to be reduced before
they are amenable to a thorough analysis.
Also, they often possess functionally important dynamic features like bistability.
In model reduction, it is sometimes more desirable to preserve the dynamic features only
than to recover a good quantitative approximation.
We present an approach to reduce the order of a bistable dynamical system
significantly while preserving bistability and the switching threshold. These properties
are important for the operation of the system in the context of a larger network.
As an application example, a bistable
model for caspase activation in apoptosis is considered.
\end{abstract}
\vspace*{-0.25cm}
\end{frontmatter}

\section{Introduction}
\label{sec:intro}

The importance of switch--like decisions in biological processes has been revealed
in a wide range of systems. Examples range from cell fate decisions via the MAP kinase cascade 
\citep{FerrellXio2001}
or apoptosis \citep{EissingCon2004} to changes of metabolic parameters, as e.g. induced
by the lac operon \citep{YildirimSan2004}.
Switch--like decisions are also a major feature of more complex dynamics, as found for oscillations in
the cell cycle \citep{PomereningSon2003}. Concerning the mathematical modeling, switch--like decisions
are usually represented by bistable dynamical systems. These systems have two stable steady states,
and depending on initial conditions or external stimuli, converge to one of these steady states.
Several theoretical approaches have been developed to study the existence of two stable steady states
in dynamical systems \citep{AngeliSon2004,EissingWal2007}.

Another important issue in the modeling of biological systems is model reduction \citep{ConzelmannRod2004}.
Due to the complexity of biological systems, typically only simplified models are amenable to a thorough
computational analysis. Methods for model reduction which are theoretically founded provide important means to
approximate a detailed description of a system by a simpler model.

Complex biological systems can often be viewed as a set of interconnected modules, each playing a
specific role. In this case, two goals for model reduction of single modules can be distinguished.
The first goal is to get a good
quantitative approximation of the original model, such that a solution of the reduced model will
differ as little as possible from a solution of the full model in the relevant variables (e.g. model
outputs). For biochemical systems, this is often achieved via a time
scale separation \citep[e.g.][]{RousselFra2001}.
The second goal focusses on preserving the qualitative dynamics of a module
and its role in the network \citep{DanoMad2006}. In large biological networks, which typically
are robust against fluctuations, the exact trajectories of the module might not be relevant,
provided its qualitative behavior is preserved such that it can maintain its role in the network.
When pursuing the latter goal, one can anticipate a much larger reduction than for the first one.
For modules having the role of switches, the goal is thus to reduce the model while preserving bistability
and the quantitative properties that are important for the module's operation, like the
switching threshold.

Based on a method introduced by \cite{SchmidtJac2004a}, which allows
to compute a measure of the contribution of each state variable in the model to bistability,
we present an approach for model reduction preserving bistability in switch--like systems.
The method is applied to a bistable system involved in programmed cell death.
We show that the reduction preserves not only bistability, but also quantitative properties like the
linear approximation to the manifold separating the two regions of attraction of the stable steady states,
which is related to the switching threshold of the system.

\section{The caspase activation model}
\label{sec:model}

We study a model for caspase activation, a major part of apoptosis, developed by
\cite{EissingCon2004}. Apoptosis, also
called programmed cell death, is a signaling program that leads the cell to commit suicide
under appropriate internal and/or external stimuli. It provides a living organism
with means to remove infected, malfunctioning or simply unneeded cells
to ensure its survival. Malfunction of apoptotic signaling
has been detected in several diseases, including developmental defects, neurodegeneration
and cancer \citep{DanialKor2004}. 
Therefore it is also of medical interest to understand the signaling network
which regulates apoptosis. 

Cell death is a switch--like decision: based on the input signal, the cell has to decide
whether to undergo apoptosis or to stay alive. There is no gradual response.
The role of the switching element in apoptosis is taken by the caspase cascade. One distinguishes initiator caspases which
receive the stimuli and effector caspases actually carrying out apoptosis.
Bistability in the model arises from a positive feedback loop between
the effector caspases and the initiator caspases, in connection with the specific inhibitors
for each type of caspases.

The model as developed by \cite{EissingCon2004}
already contains several simplifications, in particular several types of initiator and
effector caspases are combined in one species each, and the same applies to several
types of inhibitors of the effector
caspases.
Furthermore, the external stimulus is not explicitly included as an input, but instead the initial amount
of activated initiator caspases resulting from the stimulation is considered as input
to the model.

\begin{figure}
\begin{center}
\includegraphics{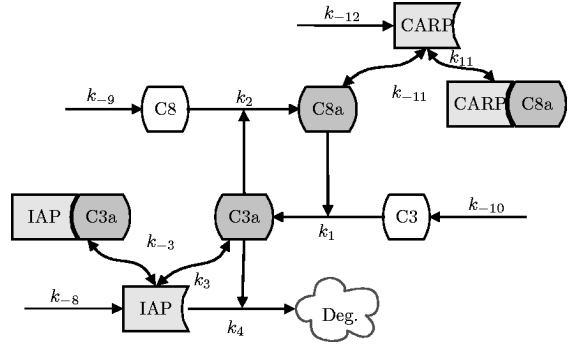}
\end{center}
\caption{Representation of the caspase activation model. Constitutive degradation reactions are not displayed.}
\label{fig:caspase-model}
\end{figure}

The species present in the model are the initiator caspase 8 and the
effector caspase 3, both in active (\texttt{C8a}, \texttt{C3a}) and inactive (\texttt{C8}, \texttt{C3}) forms, 
and the inhibitors \texttt{IAP}
and \texttt{CARP} as well as their complexes with active caspase 3 and 8, respectively.
Figure~\ref{fig:caspase-model} displays the species involved in the model and the reactions
among them.
The variables in the mathematical model denote the molecule numbers per cell of the various
species (Table~\ref{tab:model-states}).

\begin{table}[bhp]
\centering
\begin{tabular}{cc}
$
\begin{aligned}
x_1 &= [\texttt{C8}]  \\
x_2 &= [\texttt{C8a}] \\
x_3 &= [\texttt{C3}]  \\
x_4 &= [\texttt{C3a}]
\end{aligned}
$ &
$
\begin{aligned}
\quad x_5 &= [\texttt{IAP}] \\
\quad x_6 &= [\texttt{C3a\_IAP}] \\
\quad x_7 &= [\texttt{CARP}] \\
\quad x_8 &= [\texttt{C8a\_CARP}] 
\end{aligned}
$
\end{tabular}
\caption{Model state variables.}
\label{tab:model-states}
\end{table} 

The equations for the caspase activation model are given
in Table~\ref{tab:model-eq}
with parameter values as shown in Table~\ref{tab:param-values}
\citep{EissingCon2004}. These parameter values have been collected
from literature. For simplicity, we consider 
state variables (given in molecules per cell) as dimensionless.

\begin{table}
\begin{equation}
\begin{aligned}
\dot x_1 &= - k_2 x_1 x_4 - k_9 x_1 + k_{-9} \\
\dot x_2 &= k_2 x_4 x_1 - k_5 x_2 - k_{11} x_2 x_7 + k_{-11} x_8 \\
\dot x_3 &= -k_1 x_2 x_3 - k_{10} x_3 + k_{-10} \\
\dot x_4 &= k_1 x_2 x_3 - k_3 x_4 x_5 + k_{-3} x_6 - k_6 x_4 \\
\dot x_5 &= -k_3 x_4 x_5 + k_{-3} x_6 - k_4 x_4 x_5 -k_8 x_5 + k_{-8} \\
\dot x_6 &= k_3 x_4 x_5 - k_{-3} x_6 - k_7 x_6 \\
\dot x_7 &= -k_{11} x_2 x_7 + k_{-11} x_8 - k_{12} x_7 + k_{-12} \\
\dot x_8 &= k_{11} x_2 x_7 - k_{-11} x_8 - k_{13} x_8
\end{aligned}
\label{eq:fullmodel}
\end{equation}
\caption{Model equations}
\label{tab:model-eq}
\end{table}

\begin{table}[btp]
\centering
\begin{tabular}{cl@{\hspace*{0.3cm}}|@{\hspace*{0.3cm}}cl}
\hline
$k_1$ & $5.8\cdot 10^{-5}$ & $k_{11}$ & $5\cdot 10^{-4}$ \\
$k_2$ & $1\cdot 10^{-5}$ & $k_{12}$ & $1\cdot 10^{-3}$ \\
$k_3$ & $5\cdot 10^{-4}$ & $k_{13}$ & $1.16\cdot 10^{-2}$ \\
$k_4$ & $3\cdot 10^{-4}$ & $k_{-3}$ & $0.21$ \\
$k_5$ & $5.8\cdot 10^{-3}$ & $k_{-8}$ & $464$ \\
$k_6$ & $5.8\cdot 10^{-3}$ & $k_{-9}$ & $507$ \\
$k_7$ & $1.73\cdot 10^{-2}$ & $k_{-10}$ & $81.9$ \\
$k_8$ & $1.16\cdot 10^{-2}$ & $k_{-11}$ & $0.21$ \\
$k_9$ & $3.9\cdot 10^{-3}$ & $k_{-12}$ & $540$ \\
$k_{10}$ & $3.9\cdot 10^{-3}$ &  &  \\
\hline
\end{tabular}
\caption{Parameter values $[\unit{min}^{-1}]$}
\label{tab:param-values}
\end{table} 

The model yields two stable steady states, which can be identified with the \emph{living state},
where the concentrations of all active caspases---both free and bound---are equal to zero, and with
the \emph{apoptotic state} where an almost complete activation of the caspases determines the cell to undergo apoptosis.
The model captures the switch--like behavior occurring during apoptosis very well and is
thus suitable to our needs.

Furthermore, we have a \emph{threshold} manifold separating the regions of attraction, as
well as an unstable steady state on the threshold manifold which we refer to
as the \emph{decision state}, as it is relevant in the decision about the fate of
the cell.

\section{Determination of relevance}
\label{sec:relevance}

When determining the relevance of each state variable in the system to bistability, the
unstable steady state, or decision state, plays a crucial role. The boundaries of
the bistability region in parameter space are marked by bifurcations of the
unstable steady state and thus can in principle be detected by a classical bifurcation
analysis. Although bifurcation analysis reveals the influence of parameters
on the bistability, it does not give the relevance of each variable and how these are interconnected
to generate bistability.

Based on a linear approximation of the model around the decision state,
\cite{SchmidtJac2004a} perturb the influence of one state variable on the other variables
of the system. In the unperturbed case, the linear
approximation is unstable, since the decision state is unstable. 
One now searches for a perturbation that renders the decision state stable, which
would roughly correspond to reaching a bifurcation point in the original nonlinear system.
The magnitude of a perturbation found in this way is then a measure of the relevance
to bistability of the perturbed variable with its connections to other variables.

To make this approach precise, consider a system given by the differential equation
\begin{equation}
\begin{aligned}
\frac{d}{dt}x = f(x),
\end{aligned}
\label{eq:general-system}
\end{equation}
with $x\in\Real^n$.
We assume that the system is bistable and has an unstable steady state $x_d$, the decision state for the bistable behavior.
Setting $\Delta x = x - x_d$, the linearization around the decision state is given by
\begin{equation}
\begin{aligned}
\frac{d}{dt}(\Delta x) = A \Delta x,
\end{aligned}
\label{eq:linear-system}
\end{equation}
with $A = \frac{\partial f}{\partial x}(x_d)$. Note that, by our assumption, the system
\eqref{eq:linear-system} is unstable.

The linearized system is then considered as a closed feedback loop, in the sense that all
interconnections between the state variables are put into a virtual feedback path.
Breaking this feedback path yields the control system
\begin{equation*}
\begin{aligned}
L:\quad \frac{d}{dt}(\Delta x) = \tilde A \Delta x + (A-\tilde A)u,
\end{aligned}
\end{equation*}
with $u\in\Real^n$ and
\begin{equation*}
\begin{aligned}
\tilde A = \matrix{a_{11} &  & 0 \\ & \ddots & \\ 0 & & a_{nn}},
\end{aligned}
\end{equation*}
i.e. $\tilde A$ contains only the elements on the diagonal of $A$.
By setting $u=\Delta x$, the feedback path is closed again and one obtains the original linear system \eqref{eq:linear-system}.

Following \citeauthor{SchmidtJac2004a}, we assume that the matrix $\tilde A$ is stable, i.e. $a_{ii}<0$ for
$i=1,\ldots,n$. This implies that the instability of the closed loop system \eqref{eq:linear-system} is
due to interconnections among variables.

\begin{figure}
\centering
\psset{xunit=1.5cm,yunit=1.5cm,runit=1.5cm}
\begin{pspicture}(3.2,2.5)
\psset{linewidth=0.015}
\psframe(0.5,1.5)(2,2.5)
\rput(1.25,2){\scriptsize $L$}
\psline[doubleline=true]{->}(2,2)(3.2,2)
\psline[doubleline=true]{->}(2.5,2)(2.5,1.2)(0,1.2)(0,2)(0.5,2)
\psframe(0.75,0)(1.75,0.75)
\rput(1.25,0.375){\scriptsize $1+\epsilon_i$}
\psline{->}(2.5,1.2)(2.5,0.375)(1.75,0.375)
\psline{-}(0.75,0.375)(0,0.375)(0,1.2)
\uput[45](2.5,2){\scriptsize $x$}
\uput[45](2.5,0.5){\scriptsize $x_i$}
\end{pspicture}
\caption{Closed feedback loop with one feedback path perturbed}
\label{fig:feedbackloop}
\end{figure}
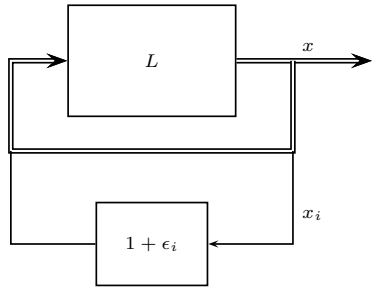 

The computation is then done as follows:
for each state variable $x_i$, a perturbation $\epsilon_i$ is introduced into the feedback path
of this variable (Fig.~\ref{fig:feedbackloop}). This yields the system
\begin{equation*}
\label{eq:perturbed-lin-sys}
\begin{aligned}
L_i(\epsilon_i):\quad \frac{d}{dt}(\Delta x) = A \Delta x + (A_i-\tilde A_i)\epsilon_i \Delta x_i,
\end{aligned}
\end{equation*}
where $A_i$ and $\tilde A_i$ are the $i$-th column of $A$ and $\tilde A$, respectively.

We are now searching for the minimal perturbation that will stabilize the system $L_i$ and define
the value $\bar\epsilon_i$ as
\begin{equation}
\begin{aligned}
\bar\epsilon_i = \min \lbrace \epsilon_i >0 \mid L_i(\epsilon_i)\text{ or }L_i(-\epsilon_i)\text{ is stable}\rbrace.
\end{aligned}
\label{eq:epsilon-computation}
\end{equation}
If the minimum does not exist, set $\bar\epsilon_i=\infty$. The higher the value of $\bar\epsilon_i$,
the more difficult it becomes to perturb the connections among the considered state variable $x_i$
and the remaining variables
such that instability is lost, and the less relevant the variable $x_i$ is to bistability.
Formally, we use the following definition of relevance.

\begin{defin}
The \emph{relevance} $R_i$ of the state variable $x_i$ to the bistability of the system \eqref{eq:general-system} is
\begin{equation*}
\begin{aligned}
R_i = \frac{1}{\bar\epsilon_i},
\end{aligned}
\end{equation*}
where $\bar\epsilon_i$ is given by equation \eqref{eq:epsilon-computation}.
\label{def:relevance-definition}
\end{defin}

\begin{figure}
\centering
\includegraphics[width=\linewidth]{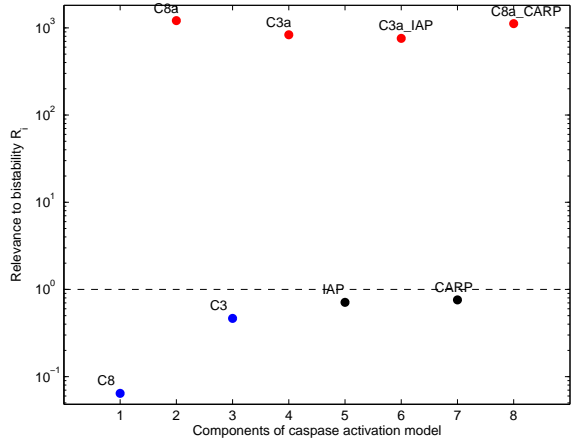}
\caption{Relevance of components to bistability in the caspase activation model}
\label{fig:comp_relevance}
\end{figure} 

This computation has been implemented numerically in the Systems Biology toolbox for Matlab \citep{SchmidtJir2005}
and has been applied to the model of caspase activation described in Section~\ref{sec:model}.
The result is shown in Fig.~\ref{fig:comp_relevance}, and one distinguishes easily the relevant components
\texttt{C8a, C3a, C3a\_IAP} and \texttt{C8a\_CARP} from the others, which are much less relevant to
bistability.

It might be possible to develop a formal approach to compute $\bar\epsilon$ in equation
\eqref{eq:epsilon-computation} based on the Kharitonov theorem and its extensions
\citep{BhattacharyyaCha1995}. However, the cost of computing a stabilizing perturbation would be similar
to the one of a direct numerical search, since we consider only one perturbation at a time. Thus this
idea is not pursued here.

\section{Bistability preserving model reduction}
\label{sec:model_reduction}

Based on the measure of relevance presented in the previous section, we now develop
a method of model reduction which preserves bistability in the reduced model. This
reduction method is then applied to the caspase activation model \eqref{eq:fullmodel}.

\subsection{Method of model reduction}
\label{ssec:model-reduction}

The basic idea of the model reduction already used by \cite{SchmidtJac2004a}
is to retain only the state variables which have been identified as
relevant to the bistability, and to neglect the dynamics of the other variables.
In contrast to \citeauthor{SchmidtJac2004a}, who replace the neglected state variables
in the remaining equations with their steady state values, we use their steady
state map which is computed from the full model. Our approach will be justified
in Section~\ref{ssec:results-interpretation}.

The relevant state variables can be identified by choosing appropriate thresholds $r>0$ for the
relevance and $\alpha>1$ which determines the gap size.
\begin{defin}
\label{def:relevant-state}
The state variable $x_i$, $i=1,\ldots,n$, is said to be \emph{relevant} 
(resp., \emph{irrelevant}) to bistability, if
there exist $r>0$ and $\alpha>1$ such that
$R_i > \alpha r$ (resp., $R_i < \frac{1}{\alpha} r$).
\end{defin}

Using this definition, the state $x$ of the system
\eqref{eq:general-system} is subdivided as
\begin{equation*}
\begin{aligned}
x \rightarrow (x_R, x_I)\T,
\end{aligned}
\end{equation*}
where $x_R$ contains the relevant variables and $x_I$ the irrelevant ones. 
Note that $\alpha$ should not be chosen
to small to get a clear distinction between relevant and irrelevant state variables.

We then proceed as follows:
\begin{enumerate}
\item For each variable which is not relevant to bistability, compute its steady state map
from the full model, i.e. solve the equation 
\begin{equation}
\begin{aligned}
0 = f(x_R, x_I)
\end{aligned}
\label{eq:implicit-steady-state}
\end{equation}
for $x_I$ to find the steady state map
\begin{equation}
\begin{aligned}
x_I = g(x_R).
\end{aligned}
\label{eq:steady-state-map}
\end{equation}

Computationally, this is similar to the division of a system in fast and
slow subsystems using a quasi-steady-state approximation \citep{SchauerHei1983}, though we do not have
fast and slow subsystems here, but rather subsystems that are relevant or irrelevant to
bistability.
The map $f$ has to be invertible with respect to $x_I$, which is the case in our example. 
Local invertibility can be checked via
the implicit function theorem, but easily checkable conditions for global invertibility are not
available in general.

\item Drop the differential equations for $x_I$
from the system
and replace the components of $x_I$ in $f_R$ by their steady state map $g(x_R)$.

The reduced model is thus given by
\begin{equation*}
\begin{aligned}
\dot x_R = f_R(x_R,g(x_R)).
\end{aligned}
\end{equation*}
\end{enumerate}

To apply the described method to the model of caspase activation~\eqref{eq:fullmodel}, 
we choose $r=10$ and $\alpha=10$. This yields the separation 
\begin{equation*}
\begin{aligned}
x_R = (x_2, x_4, x_6, x_8)\T\ \text{and}\  x_I = (x_1, x_3, x_5, x_7)\T.
\end{aligned}
\end{equation*}

For the first step in the reduction, one gets the steady
state map for $x_I$ as
\begin{equation}
\label{eq:caspase-steady-state-maps}
\begin{aligned}
x_1 &= \frac{k_{-9}}{k_9+k_2 x_4}&\quad
x_3 &= \frac{k_{-10}}{k_{10}+k_1 x_2} \\
x_5 &= \frac{k_{-8}+k_{-3} x_6}{(k_3+k_4) x_4 + k_8}&\quad
x_7 &= \frac{k_{-12}+k_{-11} x_8}{k_{12}+k_{11} x_2}
\end{aligned}
\end{equation}
and the dynamics for the reduced model are thus given by
\begin{small}
\begin{equation}
\begin{aligned}
\dot x_2 &= \frac{k_{-9}k_2 x_4}{k_9+k_2 x_4} - k_5 x_2 - \frac{(k_{-12}+k_{-11} x_8)k_{11} x_2}{k_{12}+k_{11} x_2} + k_{-11} x_8 \\
\dot x_4 &= \frac{k_{-10}k_1 x_2}{k_{10}+k_1 x_2} - \frac{(k_{-8}+k_{-3} x_6) k_3 x_4}{(k_3+k_4) x_4 + k_8} + k_{-3} x_6 - k_6 x_4 \\
\dot x_6 &= \frac{(k_{-8}+k_{-3} x_6) k_3 x_4}{(k_3+k_4) x_4 + k_8} - k_{-3} x_6 - k_7 x_6 \\
\dot x_8 &= \frac{(k_{-12}+k_{-11} x_8) k_{11} x_2}{k_{12}+k_{11} x_2} - k_{-11} x_8 - k_{13} x_8.
\end{aligned}
\end{equation}
\end{small}

Note that, in addition to the reduction in system dimension, also the dimension of the parameter space
can be reduced: 
setting $\tilde k_9 = k_9/k_2$ and $\tilde k_{10} = k_{10}/k_1$, the number of independent parameters
in the model is reduced by two.

\subsection{Results of reduction and interpretation}
\label{ssec:results-interpretation}

Analyzing the dynamics of the reduced model, one gets the following important results:
\begin{itemize}
\item All steady states of the full model are reproduced as a projection in the reduced model.
\item The steady states of the reduced model have the same local stability behavior, in particular the
same number of eigenvalues with positive real parts as their corresponding
steady states in the full model after linearization.
\item The linear approximation to the stable manifold at the decision state is the same up
to projection for both the full and the reduced model.
\end{itemize}

The first property is obtained automatically 
by keeping the steady state map \eqref{eq:caspase-steady-state-maps} of the neglected variables,
under the assumption that no two steady states are projected to the same state by the reduction.
Then one directly achieves the preservation of all steady states, which is a necessary condition
for preservation of bistability.
At this point, the change we made to the original approach of \cite{SchmidtJac2004a} is
important. Replacing the neglected variables with their constant steady state values does in general
not preserve steady states other than the unstable decision state, in particular it does not
preserve the two stable steady states in the caspase activation model.
But to preserve bistability, it is not only necessary to preserve instability of the decision state,
but one also needs to preserve the two stable steady states. The use of the steady state map provides
a general way to preserve all steady states in the reduced model, which
cannot be achieved by using constant values for the irrelevant variables.

The second and third property depend on a clear separation in relevant and irrelevant
state variables and thus on the choice of the parameters $r$ and $\alpha$.
Preservation of bistability is actually due to the second property: it guarantees that
the living steady state and the apoptotic steady state are stable in the reduced model, while the
third steady state, which is part of the threshold manifold, remains unstable.

The third property gives some quantitative measures that are reproduced exactly in the
reduced model. The stable manifold of the decision state represents the switching threshold surface
for the bistable system. 
Their linear approximations are equivalent in the original and the
reduced model. This implies that we have not only recovered the qualitative trait of the system
being bistable, but also quantitative measures like threshold values have been preserved when
staying close to the decision state. It has to be expected though that the models will differ more
when further away from the decision state, since the reduction was based on a local
analysis at the decision state.

Biologically, it is interesting that the method revealed the active forms of the caspases
as the relevant elements in generating bistability, and dropped all inactive forms and free 
inhibitors. Using
the steady state maps for these allows the reduced system to retain its role
as a biological switch in the apoptotic network. 
In contrast to a time--scale separation, the removed elements do not show a much faster
dynamics here. Yet their dynamics are not contributing to bistability, and thus can
safely be discarded. The removed variables are then considered as a pool of substances
being in quasi steady state.

\begin{figure}
\centering
\subfigure[Full model]{%
\includegraphics[width=0.9\linewidth]{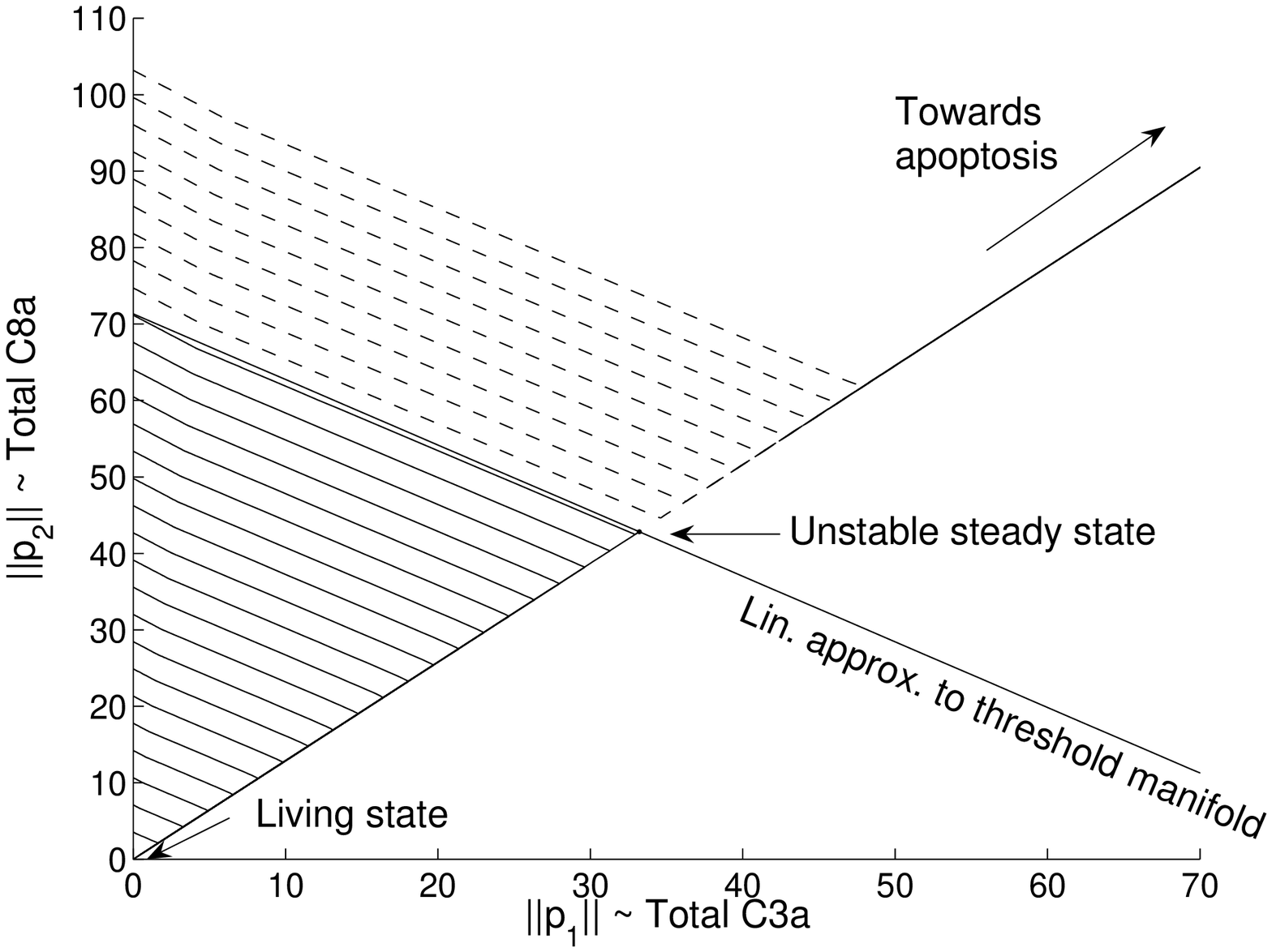}}
\\
\subfigure[Reduced model]{%
\includegraphics[width=0.9\linewidth]{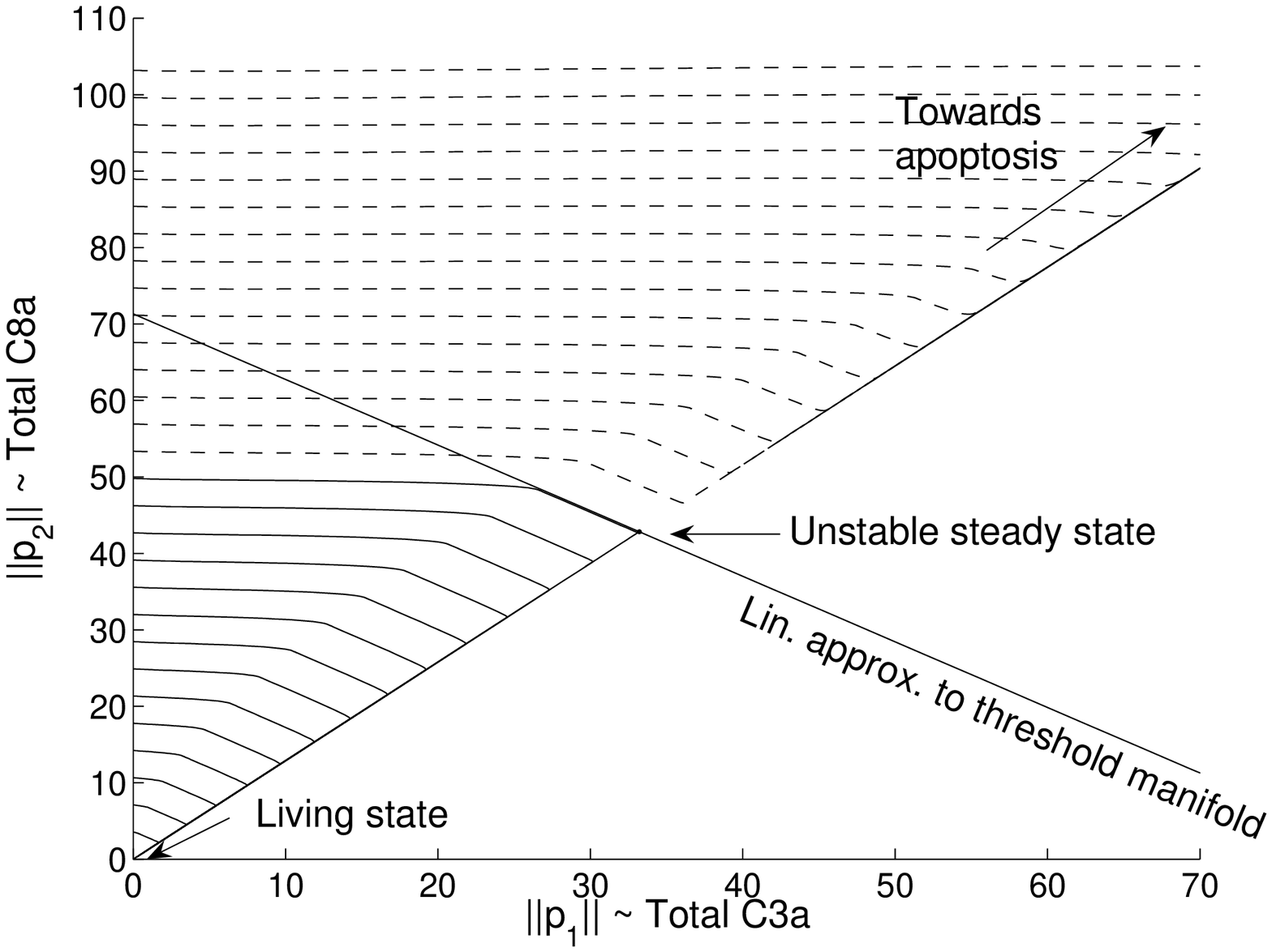}}
\caption{Trajectories of the full and the reduced model projected to a plane
spanned by vectors $p_1$ and $p_2$. These were chosen such that (a) the LATM gets
projected to a line and (b) they represent approximately the amount of total
\texttt{C3a} and \texttt{C8a}, respectively.}
\label{fig:traj_planes}
\end{figure} 

A comparison of the full and the reduced model by numerical simulation yields the results
displayed in Fig.~\ref{fig:traj_planes}, obtained by starting with different initial conditions
of free \texttt{C8a}.
This figure illustrates that the linear approximation to
the threshold manifold (LATM) is the same up to projection for both models, as well as the unstable direction of
the decision state. Furthermore, the trajectories are very similar when starting close to
the decision state.

However, we also observe that the reduced system has different dynamics when considered further away from the
decision state. In particular the threshold in initial \texttt{C8a} required to undergo apoptosis
is lower in the reduced model. Note that for the full model, the linear threshold
displayed in Fig.~\ref{fig:traj_planes} is a very good approximation of the threshold manifold,
while the approximation is not that good for the reduced model. This is likely due to the
higher nonlinearities in the reduced model introduced via the steady state maps of the neglected
components.

Although the trajectories of the full and the reduced model are different when not starting
close to one of the steady states, the results are still meeting our goals, since the focus of our
study is more on preserving bistability than recovering a quantitatively similar model.
In this way, we have identified a minimal set of components responsible for switch--like behavior.

\section{Conclusions}

We have studied a mathematical model of a
caspase activation system involved in the initiation of
apoptosis. The model is bistable to incorporate the switch--like behavior
observed in the real system. We have computed the relevance of each
state variable to bistability and observed that only four out of eight variables have a
high relevance. Using this result, the model can be reduced significantly
by retaining only the relevant state variables in the model equations plus the steady state map of the
irrelevant variables.
The reduced model retains the bistability
of the original model and also quantitative features such as trajectories close to
the steady states and the linear approximation to the threshold manifold separating the 
regions of attraction of the
two stable steady states.

\begin{small}
\bibliography{/home/waldherr/Forschung/Referenzen.bib}
\end{small}

\end{document}